\documentclass{agujournal2019}
\usepackage{url} 
\usepackage{lineno}
\usepackage{soul}
\usepackage{graphicx}
\usepackage{wrapfig}
\usepackage{array}
\usepackage{tabularx}
\usepackage{lineno}
\usepackage{xcolor}
\draftfalse

\journalname{JGR Planets}

\begin{document}

%
%


\title{Features in the visible spectra of the Enceladus particle plume and E ring: \\ Potential evidence of organic materials and/or missing sub-micron particles}
\authors{M.M. Hedman\affil{1}, S.M. MacKenzie\affil{2}}
 \affiliation{1}{Department of Physics, University of Idaho, Moscow ID 83844}
\affiliation{2}{Johns Hopkins University Applied Physics Lab, Laurel MD}
\correspondingauthor{M.M. Hedman}{mhedman@uidaho.edu}
\begin{keypoints}
\item Visible spectra of Enceladus' plume and the E ring show a change of slope around 0.5 microns.
\item This spectral feature could be due to non-icy materials in the plume particles or a deficit of sub-micron particles.
\item Variations in the plume's visible spectra could reflect variations in composition and/or typical size of the plume particles.
\end{keypoints}

\begin{abstract}
Visible spectra of the Enceladus particle plume and E ring contain evidence for a change in spectral slope around 0.5 $\mu$m.  This feature can be seen in data obtained by both the Visual and Infrared Mapping Spectrometer (VIMS) and Imaging Science Subsystem (ISS) onboard the Cassini Spacecraft, and is consistent with the slope change seen in the surface spectra of Saturn's rings and moons that has been attributed to either organics or iron compounds. The observed spectral features in the plume and E ring could represent either a non-ice contaminant in the plume particles or a deficit of sub-micron particles, so this spectral feature provides a new tool for assessing variations in the plume particle's composition and/or size distribution with time and space. The observed strength of this feature is consistent with the plume particles having an organic fraction similar to that measured by in-situ measurements, so there are good reasons to expect that this feature can be used to quantify the organic content of the plume particles. There are also hints of a potential absorption band around 0.45 $\mu$m in these spectra. If this feature can be confirmed, it could provide further constraints on the plume particles' composition. 
\end{abstract}

\section*{Plain Language Summary}
Enceladus is a small moon of Saturn that produces a plume of  water vapor and water-ice-rich particles  from a series of fissures around its south pole. Some of these erupted particles escape Enceladus' gravity and go into orbit around Saturn to form the E ring.  Visible spectra of both the Enceladus plume and E ring particles contain evidence for a change in spectral slope around 0.5 $\mu$m.  This spectral feature could represent either a non-ice contaminant in the plume particles or an absence of particles smaller than a critical size. The observed strength of this feature is consistent with the plume particles having an organic fraction similar to that measured by in-situ measurements, so this feature could {eventually} be used to quantify the organic content of the plume particles. 

\section{Introduction}

The plume of water vapor and ice emerging from Enceladus' South Polar Terrain represents one of the Cassini mission's most important scientific discoveries because it provides insights into the internal structure, composition and even habitability of icy satellites  \cite{Spencer2009, Spencer2013, Schenk2018, MacKenzie2022}. Data obtained by the various instruments onboard Cassini have already provided a great deal of information about what is going on beneath Enceladus' surface. Images show that the plume material emerges primarily from a series of four large fissures known collectively as the ``tiger stripes'' \cite{Porco2006, Porco2014, Spitale2007, Spitale2015}. Thermal infrared measurements have established that a significant amount of heat is emerging from these fissures  \cite{Spencer2006, Goguen2013}. Ultraviolet and infrared observations show that the plume material consists primarily of water vapor and small particles of water ice \cite{Hansen2006, Hedman2009}. Meanwhile, in-situ measurements have revealed that the plume's vapor component also contains molecules like H$_2$, CO$_2$ and various complex organics \cite{Waite2006, Waite2017, Waite2009}, and that the particles include varying amounts of salt, silica and various organic molecules  \cite{Postberg2011,  Postberg2018, Postberg2023, Hsu2015, Khawaja2017, Khawaja2025, Ershova2024, Nolle2024}. These compositional data place constraints on the chemistry of the subsurface system \cite{Glein2015, Postberg2023}.  Finally, gravity measurements and geological evidence suggest that a large liquid water reservoir exists beneath Enceladus' south pole \cite{Collins2007,Tobie2008, Patthoff2011, Iess2014,McKinnon2015, Thomas2016}, which may be the ultimate source of material erupting from the surface. 

In this paper we will discuss features in the visible spectra of both Enceladus' plume and the surrounding E ring (which is composed of plume particles that have escaped Enceladus and gone into orbit around Saturn) that could provide new information about the composition of the plume particles. Data obtained by the Visual and Infrared Mapping Spectrometer or VIMS \cite{Brown2004} and Imaging Science Subsystem or ISS  \cite{Porco2004} onboard the Cassini spacecraft reveal that both the plume and the E ring exhibit a change in spectral slope around 0.5-0.6 $\mu$m that is most likely due to a non-water-ice component of the plume particles. Such a feature would enable remote-sensing data to provide new constraints on the plume particles' composition that would complement the existing in-situ measurements \cite{Postberg2011, Postberg2018, Khawaja2017, Khawaja2025, Nolle2024, Ershova2024}, since trends in appropriate spectral parameters would clarify how the composition of the plume particles vary with time and source location. 

These spectral features had previously gone unnoticed because most prior investigations of the spectral properties of the plume particles have focused on near-infrared wavelengths between 1 and 5 $\mu$m  \cite{Hedman2009, Hedman2013,  Hedman2018, Dhingra2017, Sharma2023}. These spectra are dominated by the fundamental water-ice absorption band at 3 $\mu$m and broad trends that constrain the particle size distribution between 1 and 5 $\mu$m. By contrast, the relatively few published investigations of the spectral properties of Enceladus' plume at visible wavelengths either used comparisons among a small number of images obtained at extremely high phase angles by ISS to constrain the size and structure of the particles \cite{Ingersoll2011, Gao2016, Porco2017}, or noted the variations in plume's overall brightness among the observations taken with different filters must be rather subtle \cite{Ingersoll2017}.  Meanwhile, ground-based visible spectra of the E ring at low phase angles showed a clear blue slope \cite{Larson1984, Nicholson1996, dePater2004} that could be consistent with ice-rich particles, but the data were too limited to provide strong constraints on the particles' compositions \cite{Showalter1991}.   

Section~\ref{Background} below motivates searching for visible spectral features based on a combination of Enceladus' observed surface spectra and theoretical predictions for plumes with reasonable amounts of non-icy contaminants. Section~\ref{Data} then presents evidence from both ISS and VIMS observations that the spectra of both the plume and E ring contain a detectable change in spectral slope around 0.5-0.6 $\mu$m. Section~\ref{Results} then discusses these spectra and their potential implications for the composition of the plume and E ring particles. Note that for this initial study, we only perform some preliminary comparisons of the observed spectra with a limited set of model predictions. Future work will be needed to place robust quantitative constraints on the composition and size distribution of these particles.

\section{Background and Motivation}
\label{Background}

The Enceladus plume and the E ring are both faint {(with typical observed reflectances of order 10$^{-3}$ at phase angles around 160$^\circ$)} and are most detectable at high phase angles where spectral signatures are often strongly suppressed, so before conducting an in-depth analysis of the visible spectral  data it is worth motivating this work by first considering both the observed surface spectra of Enceladus and the predicted spectra of plume particles with a reasonable range of compositions. In the following subsections, we will first show that the observed visible spectra of Enceladus' surface exhibit a feature around 0.5 $\mu$m that can be attributed to a non-water-ice material, and that the spatial variations in that spectral signal are consistent with that material also being present in the plume. Next, we will generate sample theoretical spectra of the plume using Mie theory to demonstrate that non-ice materials in the plume itself could produce detectable spectral signatures at visible wavelengths and high phase angles. 

\subsection{The Visible Spectra of Enceladus' Surface}


 \begin{figure}[tbp]
 \resizebox{\textwidth}{!}{\includegraphics{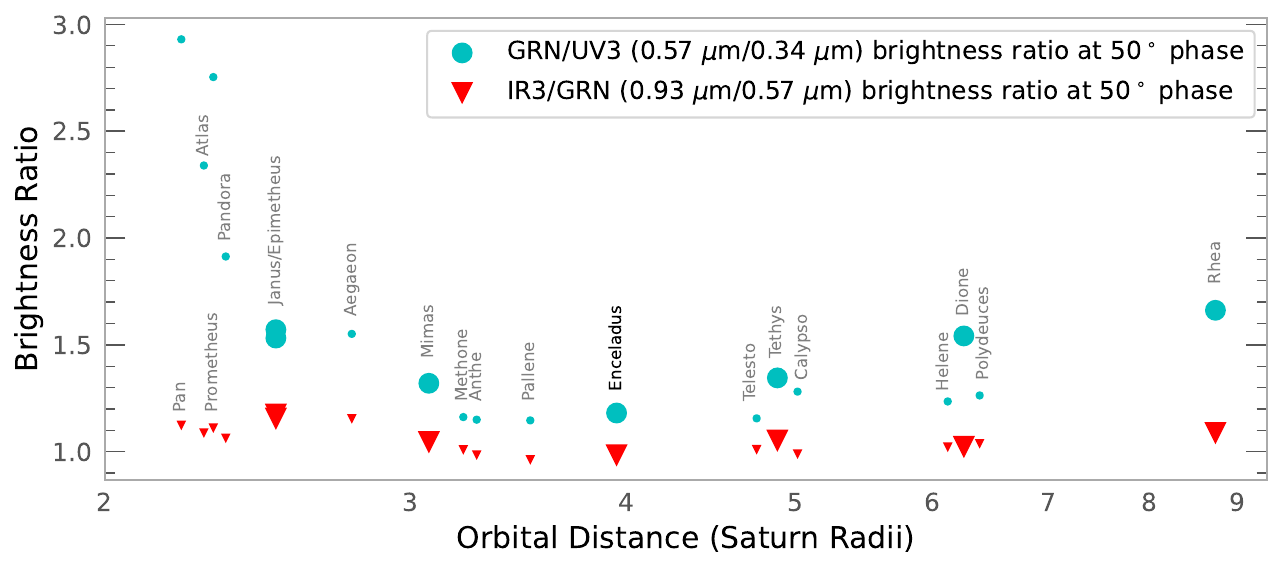}}
 \caption{Summary of the visible colors of Saturn's moons derived from Cassini images using the UV3, GRN and IR3 filters, adapted from \citeA{Ciarniello2024}. All moons show a larger GRN/UV3 color ratio than IR3/GRN color ratio, a spectral feature that is often attributed to a non-ice component of their surfaces. {The typical uncertainties in these brightness ratios is of order a few percent for the larger moons, and so is comparable to the symbol size for the larger symbols.} Note that while this spectral feature is weakest for Enceladus and the moons in its vicinity, it is still detectable even on Enceladus.}
 \label{mooncols}
 \end{figure}

While the near-infrared spectra of Saturn's rings and moons interior to Titan are dominated by features that can be attributed to very pure water ice \cite{Clark2012, Filacchione2012, Hendrix2018, Postberg2018b, Ciarniello2024}, the visible spectra of these objects are characterized by a relatively steep red slope at wavelengths shorter than 0.5 $\mu$m and a more neutral spectral slope at longer wavelengths \cite{Cuzzi2009, Clark2012,  Filacchione2012,  Thomas2018, Hendrix2018b, Ciarniello2024}.  Figure~\ref{mooncols} provides an overview of the observed colors of Saturn's moons derived from images obtained by Cassini ISS at phase angles around 50$^\circ$ \cite{Ciarniello2024}. These data demonstrate that the brightness ratio between the GRN and UV3 filters (i.e. between 0.57 and 0.34 $\mu$m) is consistently larger than the ratio between the IR3 and GRN filters (i.e. between 0.93 and 0.57 $\mu$m). This change in spectral slope, which we will generically refer to as the ``UV absorption",  is not consistent with pure ice and has been attributed to either  organic compounds or fine-grained particles rich in iron and hematite  \cite{Cuzzi2009, Clark2012, Hendrix2018, Ciarniello2024}. {The origins of this material are still not entirely clear, but it is notable that  the absorption is weakest around Enceladus' orbit, where the flux of E-ring particles is highest. E-ring particle flux is correlated with various aspects of the satellites' brightness and surface spectra  \cite{Buratti1998, Verbiscer2007, Schenk2011, Filacchione2013, Hedman2020}, and so the relative weakness of this absorption on moons with high E-ring particle fluxes suggests that its strength is probably enhanced by other processes like energetic electron bombardment \cite{Hendrix2018b}. However, while this spectral feature is weakest for Enceladus, it is still present. Since Enceladus should be covered with fresh plume deposits \cite{Kempf2010, Kempf2018, Southworth2019}, this implies that this feature could be present in the plume particles themselves.}

  \begin{figure}[tbp]
  \centerline{\resizebox{4in}{!}{\includegraphics{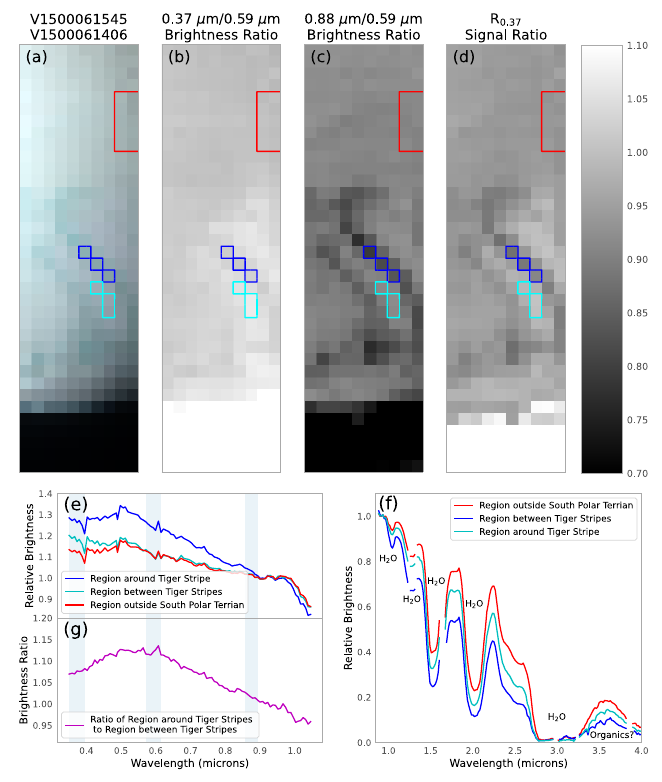}}}
 \caption{Spectra of Enceladus' surface. Panel (a) shows a composite false-color image of the region around the South Polar Terrain obtained by VIMS. In this image red, green and blue colors correspond to the brightness around 0.88, 0.59 and 0.37 $\mu$m, respectively. The tiger stripes are visible as the blue-green bands around the middle of the image. Panels (b) and (c) show brightness ratios that demonstrate that the brightness at 0.59 $\mu$m is elevated relative to the signal levels at both higher and lower wavelengths along the tiger stripes. Panel (d) shows the ratio of the observed signal at 0.37 $\mu$m to the predicted signal based on a linear interpolation of the signals between 0.88 and 0.59 $\mu$m {(defined as $R_{0.37}$ in Equation~\ref{eqr37})}. Note again how the tiger stripes appear dark in this image. Panels (e) and (f) show visible and near-infrared spectra of selected regions from the above images (Note panel (e) includes shaded bands indicating the wavelength ranges used to create panels (a)-(d)). All regions show clear water ice absorption bands between 1 and 4 $\mu$m {(the 1 $\mu$m band being responsible for the downturn at the longest wavelengths in panel (e))}, as well as a feature around 0.5 $\mu$m that is likely due to a non-ice component. Note this feature is stronger on the tiger stripes than it is in the South Polar Terrain regions between the tiger stripes. Panel (g) shows the ratio of the normalized spectra observed around a tiger stripe with a region between the tiger stripes. This ratio eliminates common structures in the spectra that are likely due to calibration artifacts and more clearly shows the structure of visible spectral feature.}
 \label{surfspec}
 \end{figure}

{Spatial variations in Enceladus' spectral properties provide further evidence that the plume particles could contain material with a detectable UV absorber. Prior work has shown that the flux of E-ring particles onto Enceladus' surface varies with latitude and longitude, and is concentrated around the active fissures within the South Polar Terrain \cite{Schenk2011, Southworth2019}.} Figure~\ref{surfspec} shows images and spectra of Enceladus derived from  two VIMS cubes (V1500061406 and V1500061545) obtained at a phase angle of around 46$^\circ$ on 2005-07-14, when the South Polar Terrain was still well illuminated. These cubes are three-dimensional datasets that provide visible and infrared spectra for each of the observed spatial pixels \cite{Brown2004}. The raw data returned by the spacecraft were calibrated using standard routines and the latest VIMS calibration pipeline designated RC19 \cite{RC19} in order to translate the raw data numbers into $I/F$, a standard measure of reflectance that is normalized to unity for a Lambertian surface viewed and illuminated at normal incidence and emission angles. 

Panel (a) of Figure~\ref{surfspec} shows a color composite image derived from these calibrated data, where the red, green and blue channels correspond to the average brightness in the wavelength ranges 0.85-0.90 $\mu$m, 0.57-0.62 $\mu$m and 0.35-0.40 $\mu$m. {Note that these three wavelength ranges differ slightly from the ones corresponding to the camera filters used in Figure~\ref{mooncols} due to a combination of VIMS' wavelength coverage and the desire to avoid wavelength channels that could be become saturated}. For the sake of brevity, we designate the observed signals in each of these three wavelength bands as $S_{0.88}$, $S_{0.59}$ and $S_{0.37}$, respectively.  The most obvious features in the color image are two blue lineations that correspond to the tiger stripes Damascus and Baghdad (compare with the IR images shown in \citeA{Brown2006}), which are also clearly blue in the visible imaging data \cite{Schenk2011}.  Panels (b) and (c) of Figure~\ref{surfspec} show the brightness ratios $S_{0.37}/S_{0.59}$ and $S_{0.88}/S_{0.59}$, and in both of these panels Baghdad {appears as a dark line running diagonally across the South Polar Terrain}. The low value of $S_{0.88}/S_{0.59}$ is consistent with the generally blue color of the tiger stripes, but the lower value of $S_{0.37}/S_{0.59}$ indicates that the tiger stripes are actually less blue than their surroundings at the shortest wavelengths. This would imply that the UV absorption is stronger in the tiger stripes than it is in the surrounding terrain. 

Variations in the UV absorption can also be documented using ratio of the observed signal $S_{0.37}$ to the predicted signal one would see at 0.37 $\mu$m based on a linear extrapolation of $S_{0.59}$ and $S_{0.88}$. This ratio is given by the following expression:
\begin{equation}
R_{0.37}=\frac{(\lambda_{0.88}-\lambda_{0.59})S_{0.37}}
{(\lambda_{0.88}-\lambda_{0.37})S_{0.59}-(\lambda_{0.55}-\lambda_{0.37})S_{0.88}}
\label{eqr37}
\end{equation}
where $\lambda_{0.37}$, $\lambda_{0.59}$ and $\lambda_{0.88}$ are the appropriate average wavelengths. Panel (d) of Figure~\ref{surfspec} shows how this ratio varies across Enceladus' surface. Interestingly, both Baghdad and Damascus appear as dark lines in this image, which implies a stronger UV absorption along the tiger stripes. By contrast, the regions between the tiger stripes appear bright in these images, indicating a weaker UV absorption in this portion of the South Polar Terrain. 

Panels (e) and (f) of Figure~\ref{surfspec} show average spectra of representative regions indicated with colored lines in Panels (a)-(d). These regions were selected to avoid regions where VIMS was saturated at any wavelengths, and are all normalized to unity between wavelengths of 0.90 and 0.95 $\mu$m in order to facilitate comparisons in the spectral shape.  At wavelengths  above 1 $\mu$m the spectra are clearly dominated by the standard water ice bands at  1.04, 1.25. 1.5, 2.0 and 3.0 $\mu$m. The  spectra of the regions around and between the  tiger stripes also show some weak dips between 3.4 $\mu$m and 3.6 $\mu$m that might be due to organic compounds \cite{Brown2006}. Turning to the visible spectra, the continuum slope between 0.5 and 0.9 $\mu$m  is clearly much bluer around the tiger stripes than it is elsewhere on the {moon} \cite{Schenk2011}, a trend that has been interpreted as evidence for larger effective grain sizes in the regolith closer to these fissures \cite{Jaumann2008}. However, these spectra also become noticeably shallower between 0.3 and 0.5 $\mu$m, which is likely due to the UV absorption. 
 
One complication with interpreting these spectra is that all the spectra contain sharp features around 0.4, 0.5 and 0.6 $\mu$m. Since these features have the same shape and amplitude in all the spectra, they are likely subtle calibration artifacts rather than real features in the spectra. Note that these structures correspond to brightness variations of order 5\%, and so are subtle compared with the correction factors introduced in the latest version of the calibration pipeline \cite{RC19}. Fortunately, we can still document variations in the strength of the UV absorption by comparing the signals from the different regions. On the one hand, the spectra of the region between the tiger stripes (cyan) and the spectra of the region outside the Solar Polar Terrain (red) are nearly identical between 0.6 and 1.0 $\mu$m, but the two spectra diverge at shorter wavelengths starting around 0.6 $\mu$m, with the region between the tiger stripes becoming 5\% brighter than the region outside the South Polar Terrain at 0.35 $\mu$m. This is consistent with the regions between the tiger stripes having a weaker UV absorption, as shown in Panel (d). More importantly, Panel (g) shows the ratio of the normalized spectrum of the region around Baghdad Sulcus (blue in panels e and f) to the spectrum of the region between the tiger stripes. This ratio shows a clear blue slope at wavelengths longer than 0.6 $\mu$m that gradually turns over between 0.5 and 0.6 $\mu$m to a red slope below 0.5 $\mu$m. This is consistent with the shape of the UV absorption seen elsewhere in the Saturn system \cite[Figure~\ref{mooncols}]{Clark2012, Filacchione2012, Hendrix2018, Postberg2018, Ciarniello2024}, and  confirms that the UV absorption is strongest in the regions around the tiger stripes.

The VIMS surface spectra indicates that the non-water-ice material responsible for the UV absorption is concentrated near the plume sources, but also that this material is relatively rare in the regions between the fissures. Since the entire South Polar Terrain is expected to be covered in thick deposits of plume material \cite{Southworth2019}, these findings might at first appear to make contradictory predictions about whether plume material should contain the UV absorber. However, recent analysis of the in-situ measurements of the plume particle composition by the Cosmic Dust Analyzer (CDA) by \citeA{Ershova2024} could clarify this situation. That work found that particles with different compositions probably had very different dynamics, with the particles launched in well-collimated jets being more rich in organic compounds, while the particles launched over a wider range of angles from the vents could be more salt-rich \cite{Ershova2024}. Since organic materials are expected to have a stronger UV absorption than salts, this model could be consistent with the above observations if the deposits between the tiger stripes contained a larger fraction of the salt-rich plume material and the regions immediately around the tiger stripes contained more organic-rich material. In this context, it is reasonable to expect that the plume itself could contain some amount of the material that produces the UV absorption.

\subsection{Predicted visible spectral signatures of plume particles.}

Even if the plume material contains the non-icy material responsible for the UV absorption in the surface deposits, it is not immediately obvious whether this signal could be detectable in the spectra of the plume itself. Typical surface spectra are obtained at low phase angles, but the particle plume itself can only be clearly detected at high phase angles. In this viewing geometry, the observed signal is primarily due to diffraction around individual plume particles, which tends to suppress absorption bands. Indeed, infrared spectra of the plume only clearly show the fundamental 3-$\mu$m ice band, and the weaker overtone bands at 1.5 $\mu$m and 2.0 $\mu$m are largely absent \cite{Hedman2009, Dhingra2017, Sharma2023}. 

Fortunately, it is relatively straightforward to calculate how the plume's particle composition and size distribution should affect its spectral properties because it is a very tenuous system. Multiple scattering among particles can therefore be neglected and the observed spectrum of the plume is simply the sum of the light scattered by the individual particles \cite{Porco2006, Hedman2009, Ingersoll2011}.  Furthermore, the observations where the plume is detectable are all obtained at high phase angles, where the scattering efficiency of a particle with a given size and optical constants/composition can be reasonably  well approximated using Mie scattering theory \cite{Hedman2009, Ingersoll2011, Gao2016, Sharma2023}. 

\begin{figure}[tb]
\resizebox{\textwidth}{!}{|\includegraphics{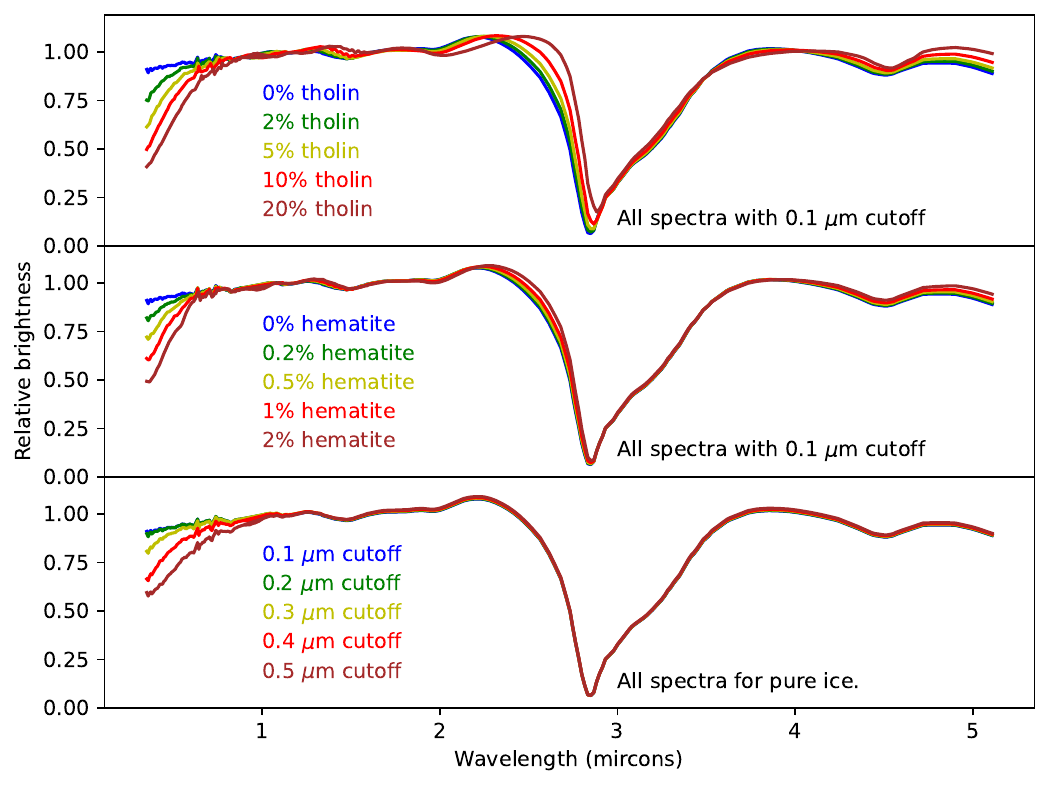}}
\caption{Model spectra of particle populations observed at a phase angle of 160$^\circ$, computed using Mie Theory. All particle populations are assumed to follow a power-law size distribution with differential index of -2.5. The top panel shows spectra of water-ice particles assuming optical constants from \citeA{Warren2008} with different concentrations of \citeA{Baratta2015} tholins, the middle panel shows spectra of ice particles with different concentrations of hematite \cite{Querry1985}, and  the bottom panel shows the spectra of pure ice particles with various values for the minimum particle size.}
\label{theory}
\end{figure}

Figure~\ref{theory} shows the predicted plume spectra across the entire VIMS wavelength range for a variety of assumptions about the plume particle properties, computed {using the {\tt SF\_SD} function from the } PyMieScatt software package \cite{Sumlin2025}. These spectra are computed assuming the plume is observed at a phase angle of 160$^\circ$ and that the particles in the plume follow a power-law size distribution with a power-law index of -2.5, which yields an infrared spectrum that is roughly consistent with those observed for the plume \cite{Hedman2009, Dhingra2017,Sharma2023}. 

For the spectra in the top panel of Figure~\ref{theory} the power-law size distribution is truncated at a minimum size (radius) of 0.1 $\mu$m and a maximum size of 5 $\mu$m, and the particles are assumed to be composed primarily of water ice with varying amounts of tholins. In practice, these spectra are computed using the optical constants for water ice from \citeA{Warren2008} and the {{optical constants for tholins provided in {\tt https://ocdb.smce.nasa.gov} derived from \citeA{Baratta2015} (Spectra computed using tholin optical constants from \citeA{Khare1984} were almost identical with these spectra at this scale).} and we compute the optical constants of the mixture using the Maxwell-Garnett rule from Effective Medium Theory:
\begin{equation}
\epsilon_{mix}=\epsilon_{w}\left[1+ \frac{3f_t\frac{\epsilon_t-\epsilon_w}{\epsilon_t+2\epsilon_w}}{1-f_t\frac{\epsilon_t-\epsilon_w}{\epsilon_t+2\epsilon_w}}\right]
\end{equation}
where $\epsilon_w$ and $\epsilon_t$ are the assumed complex dielectric constants for the water ice and the tholin, and $f_t$ is the assumed volume fraction of tholin in the mixture \cite{BH1983}. Figure~\ref{theory} shows that even only a few percent of tholins in the ice grains could have a detectable effect on the visible spectrum of the plume particles. 

While the percentage of tholin-like organics in the plume particles is still uncertain, Cassini's CDA has measured the composition of particles in both the plume and the E ring, and found that while most of the particles are nearly pure ice, there are substantial fractions of particles that contain significant amounts of salt or organics \cite{Postberg2018b, Khawaja2025}. In particular, CDA measurements of the E-ring particles indicate that about 25\% of the E-ring grains and plume grains are organic-rich, although only about 2\% of the E ring-grains contain high-mass organic compounds that would be most analogous to tholins \cite{Postberg2018b, Nolle2024}. Furthermore, CDA measurements within the plume itself could indicate the organic-rich particles are more common at low altitudes \cite{Khawaja2017}, although more recent modeling has complicated this picture by suggesting that organic particles might be launched at higher speeds, and some of the particles that were thought to be pure ice might in fact contain some amount of organics \cite{Ershova2024}. With these uncertainties in mind, these calculations indicate that the observed organic component of the plume material could have a detectable effect on the plume's visible spectrum, even at high phase angles. 

It is important to note that this sort of UV absorption is not a unique signature of complex organics like those found in tholins. {In particular, previous studies have proposed that iron-containing compounds like hematite could provide an alternative explanation for the UV absorption in the Saturn system \cite{Cuzzi2009, Clark2012}.}
The middle panel of Figure~\ref{theory} shows spectra of a plume whose particles are composed of mixtures of water ice and hematite, assuming the optical constants from \citeA{Querry1985}. Hematite produces an absorption at short visible wavelengths at even smaller concentrations than tholins, owing to its larger imaginary refractive index at these wavelengths. CDA measurements do not yet provide strong constraints on the iron content of the plume particles due to a nearby water-ion signal \cite{Postberg2008},  and so we cannot rule out the possibility that hematite could be a significant constituent of the Enceladus plume.  {While iron-containing compounds could be  involved in the water-rock interactions in Enceladus' internal ocean \cite{Glein2018, Ramirez2024}, it is not yet clear whether detectable amounts of these materials  could become incorporated into plume particles. Since in-situ measurements provide more concrete predictions about the presence of organics in the plume particles \cite{Postberg2018b, Nolle2024}, we will focus primarily on tholins in this paper.}

Another possible way to produce a reduction in the plume's brightness at short wavelengths is an abrupt cut-off in the particle size distribution. Small ice-rich particles viewed at high phase angles scatter most efficiently at wavelengths comparable to the particle sizes \cite{Hedman2009, Sharma2023}, so a deficit of sub-micron particles would naturally reduce the plume's brightness at short visible wavelengths.  The bottom panel of Figure~\ref{theory} demonstrates this by showing example spectra for populations of pure ice particles with power-law size distributions with different minimum particle sizes. Increasing the value of the minimum particle size produces a stronger red slope at short wavelengths while leaving the rest of the spectrum largely unchanged.  However, such a sudden reduction in the slope of the particle size distribution for the plume would be inconsistent with in-situ measurements. Comparisons of particle densities measured by the Cassini Plasma Spectrometer (CAPS), Cosmic Dust Analyzer (CDA), and Radio and Plasma Wave Spectrometer (RPWS) instruments are consistent with an approximately power-law size distribution extending from just a few nanometers to a few microns \cite{Dong2015}. While these measurements allow for some variations in the power-law index between 0.1 $\mu$m and 1 $\mu$m, those sorts of features in the size distribution are unlikely to produce features in the plume spectra comparable to those from tholins at concentrations of order a few percent. The sub-micron particle size distribution for the broader E ring is less certain. Some in-situ data could indicate that the E-ring's particle size distribution follows a power law into the sub-micron size range \cite{Ye2014, Ye2016, Ye2018}. However, spectral and photometric properties of the E ring suggest that the particle size distribution peaked around 1 $\mu$m \cite{Showalter1991}, which could be consistent with dynamical simulations that indicate sub-micron particles have shorter lifetimes within the E ring \cite{Juhasz2007}. 

{It is also worth noting that all the high-phase spectra shown in Figure~\ref{theory} exhibit a red spectral slope at wavelengths shorter than 0.5 $\mu$m, which is in stark contrast with the strong blue slope observed in E-ring spectra obtained at low phase angles by Earth-based telescopes \cite{Larson1984, Showalter1991, Nicholson1996, dePater2004}.  While it is true that adding either tholins and hematite to the E ring should cause the spectral slope to become more red at both low and high phase angles, it turns out that low-phase spectra of dusty systems are more difficult to accurately predict and model than high-phase spectra. For one, the observed signal at high phase angles is primarily due to diffraction around particles, which means that at a particular phase angle the signal at any given wavelength is most efficiently scattered by particles with a limited range of sizes \cite{Hedman2009}. By contrast, at low phase angles particles with a wider range of sizes can potentially scatter light at each wavelength, complicating efforts to translate the observed spectrum into information about the particle size distribution \cite{Showalter1991}. Furthermore, the observed brightness of particle populations becomes more sensitive to the shapes of the individual particles at low phase angles, making Mie theory much less accurate  \cite{PollackCuzzi1980}. Indeed, preliminary Mie-theory-based calculations for particle populations following power-law size distributions failed to yield predicted spectra  that reproduced the low-phase E-ring spectra. We therefore conclude that reliably constraining the particles' composition and size distribution with the low-phase spectra is not straightforward, and the blue slope of these spectra does not necessarily rule out the possibility that the high-phase E-ring spectra exhibit a UV absorption.}

In summary, the theoretical calculations  shown in Figure~\ref{theory} demonstrate that complex organic compounds like tholins could have detectable effects on the observed visible spectra of the Enceladus plume and the E ring, so long as the concentrations of these particles are order a few percent. Similar spectral features could also potentially be produced by small amounts of hematite or by strong deficits of sub-micron particles.

\section{Evidence for a visible spectral feature in the plume and E ring}
\label{Data}

Evidence of a UV absorption can be found in spectra of both the plume and the E ring obtained by both ISS and VIMS. These data also suggest that the strength of this spectral feature in the plume may vary with both space and time.

\begin{table}
\caption{ISS images considered in this study.}
\label{imtab}
\resizebox{\textwidth}{!}{{\begin{tabular}{|c|c|c|c|c|c|c|c|c|}\hline
Image Name & Observation & Filter & Wavelength & Range$^a$ & W Long. & Phase  & Orbital Phase$^b$ & offsets$^c$ \\
& Start Time & & (microns) &  (km) & (deg) & (deg) & (deg) & (pixels) \\  \hline
N1671552882 & 2010-12-20T15:26:59 & IR3 & 0.928 & 156063 & 223.9 & 163.4  & 357.3 & 5,60\\
N1671552992 & 2010-12-20T15:29:16 & IR1 & 0.750 & 155609 & 224.4 & 163.2 & 357.7 & 0,20 \\
N1671553072 &2010-12-20T15:30:37 &  RED & 0.649 & 155315 & 224.8 & 163.1 & 357.9 & -1,30  \\
N1671553214 & 2010-12-20T15:32:41 & BL1 & 0.455 & 154833 & 225.4 & 162.9 & 358.3 & -2,18 \\
N1671553594 & 2010-12-20T15:37:23 & UV3 & 0.343 & 153662 & 226.8 &  162.5 & 359.2 & 0,0 \\
\hline
\end{tabular}}}

$^a$ Distance between the spacecraft and Enceladus' center.

$^b$ Angle between Enceladus' orbital longitude and its orbital pericenter.

$^c$ Pixel offsets used to align images after re-scaling by range ratios.

\end{table}

\begin{figure}[tbp]
\resizebox{\textwidth}{!}{\includegraphics{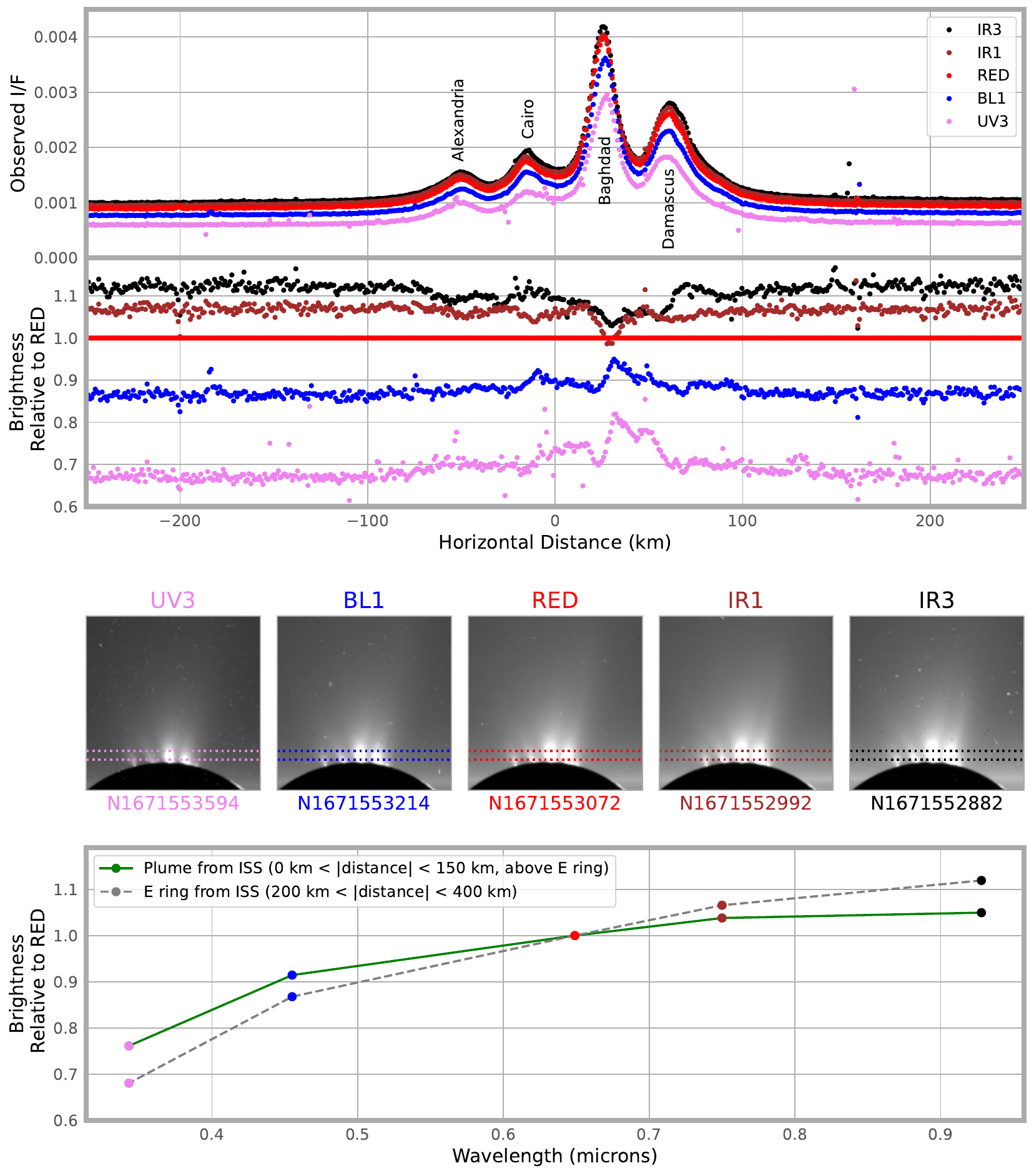}}
 \caption{Spectra of the Enceladus plume and E ring from multi-wavelength ISS images. The five panels show cropped and re-scaled versions of images of the plume {(corresponding to pixels in columns 310-710 and rows 610-1010 for the highest-resolution UV3 image)}, where material can be seen erupted from Alexandria, Cairo, Baghdad and Damascus sulci from left to right. All images use a common stretch and so clearly show the plume is fainter in the UV3 and BL1 filters than it is at longer wavelengths. The top panels show the average horizontal brightness profile for the regions indicated by the dotted lines in the images, as well as the brightness relative to the values observed with the RED filter. {The horizontal distances in these plots are measured relative to the point where the region gets closest to Enceladus' surface and computed based on the range to Enceladus' center.} The bottom panel shows the spectrum of the E ring (corresponding to the average signal between horizontal distances of $\pm$200 km and $\pm$400 km), and the plume (corresponding to the average signal above the E-ring background signal within horizontal distances of $\pm$150 km).}
 \label{plumeiss}
 \end{figure}

\subsection{Evidence for a UV absorption from multi-wavelength ISS images}
  
Cassini ISS obtained low-resolution spectra of the plume and E ring by observing the material around Enceladus through different filters. For this initial study, we focus on a set of five narrow-angle camera images that were obtained over the course of about 12 minutes in late 2010 using five different filters  spanning the wavelength range 0.34-0.93 $\mu$m. This particular sequence was chosen because it provided the cleanest signals from the E ring and the plume {based on the several considerations.} Most importantly, the moon's limb was  mostly dark, which eliminated the need to consider stray light from a lit crescent. In addition, the images were obtained from a geometry where the signals from the individual fissures are clear, and at a range that was close enough for those different sources to be clearly identifiable but also large enough that the changes in viewing geometry among the different images are small (see Table~\ref{imtab}). 

These images were calibrated using the standard calibration routines implemented on the PDS ring-moon system node \cite{West2010, Knowles2020}. Each image was then re-scaled based on the ranges to Enceladus center provided in Table~\ref{imtab}, leaving the last and highest-resolution image unchanged. We then aligned the images based on the position of Enceladus' dark disk (re-scaled pixel offsets for each image are provided in Table~\ref{imtab}). Finally, we subtracted the median signal {of} a region on the dark side of Enceladus from each re-scaled image in order to remove any instrumental background levels (the specific region used for this purpose corresponds {to} all the pixels in columns 500-600 and rows 1000-1020 of the rescaled images). Figure~\ref{plumeiss} show cropped versions of the resulting images using a common linear stretch. Material can clearly be seen erupting from the four fissures (Alexandria, Cairo, Baghdad and Damascus from left to right), and the entire scene behind Enceladus is clearly brighter than the dark side of Enceladus due to the background E ring. Note that both the plume and the E ring are fainter in the images obtained through the short-wavelength UV3 and BL1 filters than they are in the three longer-wavelength filters (RED, IR1 and IR3), which suggests that both these dust populations show evidence of the UV absorption. 

In order to quantify these spectral trends, we compute a horizontal brightness profile from each rescaled, aligned, and background-subtracted image as the average brightness in pixel rows 960-980 (that is, the region between the dotted lines shown in Figure~\ref{plumeiss}). {This specific region was chosen to minimize  imaging artifacts due to cosmic rays (which appear as bright dots in the images), and because the horizontal region remains above the G ring (which appears as a narrow bright band near the bottom of each image), so the dominant signal comes from the E ring at large horizontal distances.} The top panel of Figure~\ref{plumeiss} shows the resulting brightness profiles, which show signals from all four tiger stripes above a baseline offset level due to the E ring.  In order to better quantify the differences between these profiles, we also plot the ratio of each profile relative to the profile derived from the image obtained in the RED filter. Together, these profiles demonstrate that the spectra of both the E ring and the plume have a clear red spectral slope. The E ring shows consistent brightness ratios, with the IR3  data being about 10\% brighter than the RED data, and UV3 data being about 30\% fainter than the RED data. Within the plume, the brightness in the different filters converge somewhat, but even in the core of the plume emerging from Baghdad (where the signal is 2-4 times the background signal from the E ring), the UV3 signal remains 20\% fainter than the RED, IR1 and IR3 signals. {Note that variations in the altitudes probed along these horizontal profiles can complicate their detailed interpretation, but the general properties of the color ratios do not appear to be extremely sensitive to the selected region.}

Interestingly, the variations in the brightness ratios are not the same shape as the variations in the plume's overall brightness. In particular, the largest UV3/RED ratios and the smallest IR3/RED ratios are found in the region between the peak signals from Baghdad and Damascus. This suggests that the composition {and/or size distribution} of the plume particles may vary among the different sources in nontrivial ways. This certainly  merits further investigation, but such analysis will need to carefully consider both the changes in viewing geometry \cite{Porco2014, Spitale2015} and short-time-scale variations in the activity of individual sources \cite{Ingersoll2020, Spitale2025}. 

Given the above challenges with quantifying the spatial variations in the plume spectra, we will focus here on the average spectra of the plume and E ring derived from the horizontal brightness profiles shown in the top panel of Figure~\ref{plumeiss}. The E-ring spectrum is computed by taking the average brightness in these profiles between horizontal distances of 200 and 400 km on either side of zero. Meanwhile, the plume spectrum is constructed by taking these same horizontal profiles, removing the average E-ring spectrum, and then computing the average remaining brightness within 150 km horizontal distance from 0. The resulting brightness values are provided in Table~\ref{imtabspec} and the normalized spectra are shown in the bottom panel of Figure~\ref{plumeiss}. Note that due to the large number of pixels included in these averages, the statistical uncertainties in these brightness estimates are very small, which means the uncertainties in these spectra are dominated by calibration uncertainties, which are included in Table~\ref{imtabspec} and range between 3\% and 6\%. Both these spectra show evidence of an UV absorption, which will be discussed in more detail below.

\begin{table}
\caption{Average spectra of the E ring and Plume}
\label{imtabspec}
\begin{tabular}{|c|c|c|c|c|}\hline
Filter & Wavelength$^a$ & Calibration & E ring & Plume \\
  & (microns) & Uncertainty$^b$  & (I/F) & (I/F) \\
  \hline
IR3 & 0.928 & 0.0621 & 0.001020 & 0.000553 \\
IR1 & 0.750 & 0.0404 & 0.000971 & 0.000547 \\
RED & 0.649 & 0.0289 & 0.000911 & 0.000527 \\
BL1 & 0.455 & 0.0303 & 0.000791 & 0.000482 \\
UV3 & 0.343 & 0.0647 & 0.000620 & 0.000401 \\
\hline
\end{tabular}

$^a$ from \citeA{Porco2004}

$^b$ Fractional uncertainty in calibration from \citeA{Knowles2020}

\end{table}

\subsection{Evidence for a UV absorption from high signal-to-noise VIMS spectra}

The VIS channel of the VIMS instrument provides three-dimensional data cubes consisting of 96 images at a series of wavelengths between 0.35 and 1.05 $\mu$m. These data can provide more detailed information about the shape of the E-ring and plume spectra. However, this instrument also requires longer integration times to obtain sufficient signal-to-noise on faint targets like the plume and the E ring, and has much lower spatial resolution, which complicates efforts to isolate the plume and E ring signals. We therefore elected to initially focus only on the same sequence of observations obtained in late 2005 that yielded the IR spectra analyzed by \citeA{Hedman2009}. Due to the unique geometry of this particular encounter with Enceladus, this set of observations occurred during one of the longest periods of time when the plume was observed at high phase angles, and so provides the best opportunity for high signal-to-noise spectra. Furthermore, these observations also used the higher-resolution HIRES mode for the VIS channel \cite{Brown2004, Filacchione2007}, which facilitates detection of the plume. The main complication with this early observation is that the limb immediately below the plume is lit, which needs to be accounted for in extracting the plume signal. 

\begin{table}
\caption{VIMS cubes used in this study}
\label{vimstab}
\resizebox{\textwidth}{!}{\begin{tabular}{|c|c|c|c|c|c|c|c|c|} \hline
Observation & Filename &  Image Midtime & Image Size & Exposure & Range & W Long. &  Phase & Orb. Phase \\
& & & (pixels) & Time VIS  (s) & (km) & (deg) & (deg) & (deg) \\
\hline
003 & V1511793703 & 2005-331T14:14 & 40x30 & 7.2 & 126,220 &  105.0 & 160.6 & 100.4 \\
003 & V1511794087 & 2005-331T14:26 & 40x30 & 16 & 126,874 & 105.7 & 160.7 & 102.5 \\
003 & V1511794976 & 2005-331T14:39 & 40x30 &19 & 128,375 & 107.2 & 160.9& 104.8 \\
003 & V1511795691 &2005-331T14:48 & 40x30 & 7.2 & 129,570 & 108.6 & 161.1& 106.4 \\
003 & V1511795992 & 2005-331T14:56 & 40x30 & 19 & 130,071 & 109.2 & 161.1 & 107.9 \\
003 & V1511796659 & 2005-331T15:09 & 40x30 & 16 & 131,177 &110.6 & 161.2 & 110.2 \\
003 & V1511797548 & 2005-331T15:20 & 40x30 & 13 & 132,647 & 112.5 & 161.3 & 112.3 \\
003 & V1511798376 & 2005-331T15:38 & 40x30& 16 & 134,019 & 114.5 & 161.3 & 115.4 \\ 
003 & V1511799464 & 2005-331T15:52 & 40x30 & 11 & 135,843 & 117.2 & 161.3 & 118.0\\
003 & V1511799891 & 2005-331T15:58 & 40x30 & 7.2 & 136,569 & 118.3 & 161.3 & 119.1 \\
\hline
010 & V1511800181 & 2005-331T16:05 & 27x27& 17 & 137,068 & 119.1 & 161.3 & 120.4 \\
010 & V1511800741 & 2005-331T16:14 & 27x27& 17 & 138,045 & 120.6 & 161.3& 122.1 \\ 
\hline
001 & V1511801493 & 2005-331T16:28 & 30x30& 19 & 139,231 & 122.5 & 161.3 & 124.6 \\ 
001 & V1511802247 & 2005-331T16:40 & 30x30& 19 & 140,794 & 125.0 & 161.3 & 126.8 \\
001 & V1511803001 & 2005-331T16:53 & 30x30& 19 & 142,262 & 127.3 & 161.3 & 129.1 \\ 
\hline
011 & V1511807379 & 2005-331T18:04 & 30x30& 12 & 152,790 & 142.0 & 161.4 & 142.0 \\ 
011 & V1511807805 & 2005-331T18:13 & 30x30& 19 & 156,213 & 143.8 & 161.4 & 143.6 \\ 
\hline
007 & V1511808707 & 2005-331T18:25 & 24x30& 7.7 & 156,947 & 147.1 & 161.3 & 145.8 \\
007 & V1511809910 & 2005-331T18:46 & 30x30& 12 & 161,222 & 151.6 & 161.2 & 149.6 \\
007 & V1511810387 & 2005-331T18:51 & 24x30& 9.6 & 163,095 & 153.5 & 161.2 & 150.9 \\ 
\hline
\end{tabular}}
\end{table}

\begin{table}
\caption{Parameters used in processing VIMS cubes}
\label{vimstab2}
\begin{tabular}{|c|c|c|c|c|c|} \hline
Observation & E-ring  & E-ring  & Plume & Plume & Enceladus \\
&  signal region & background & signal & background & nom. center \\
&  (rows,columns) & (rows,columns) & (rows) & (rows) & (column pixel) \\
\hline
003 & 0-3,15-24 & 10-19,15-24 & 12-21 & 8-11,22-26 & 19 \\
010 & 0-3,10-19 & 10-19,10-19 &12-21 & 8-11,22-26 & 15.5 \\
001 & 0-3,12-21 & 10-19,12-21 & 11-20 & 7-10,21-25 & 17 \\
011 & 0-3,12-21 & 10-19,12-21 & 11-20 & 7-10,21-25 & 16.5 \\
007 & 0-3,11-20 & 10-19,11-20 & 11-20 & 7-10,21-25 & 14 \\
\hline
\end{tabular}
\end{table}

\begin{figure}[tbp]
\resizebox{\textwidth}{!}{\includegraphics{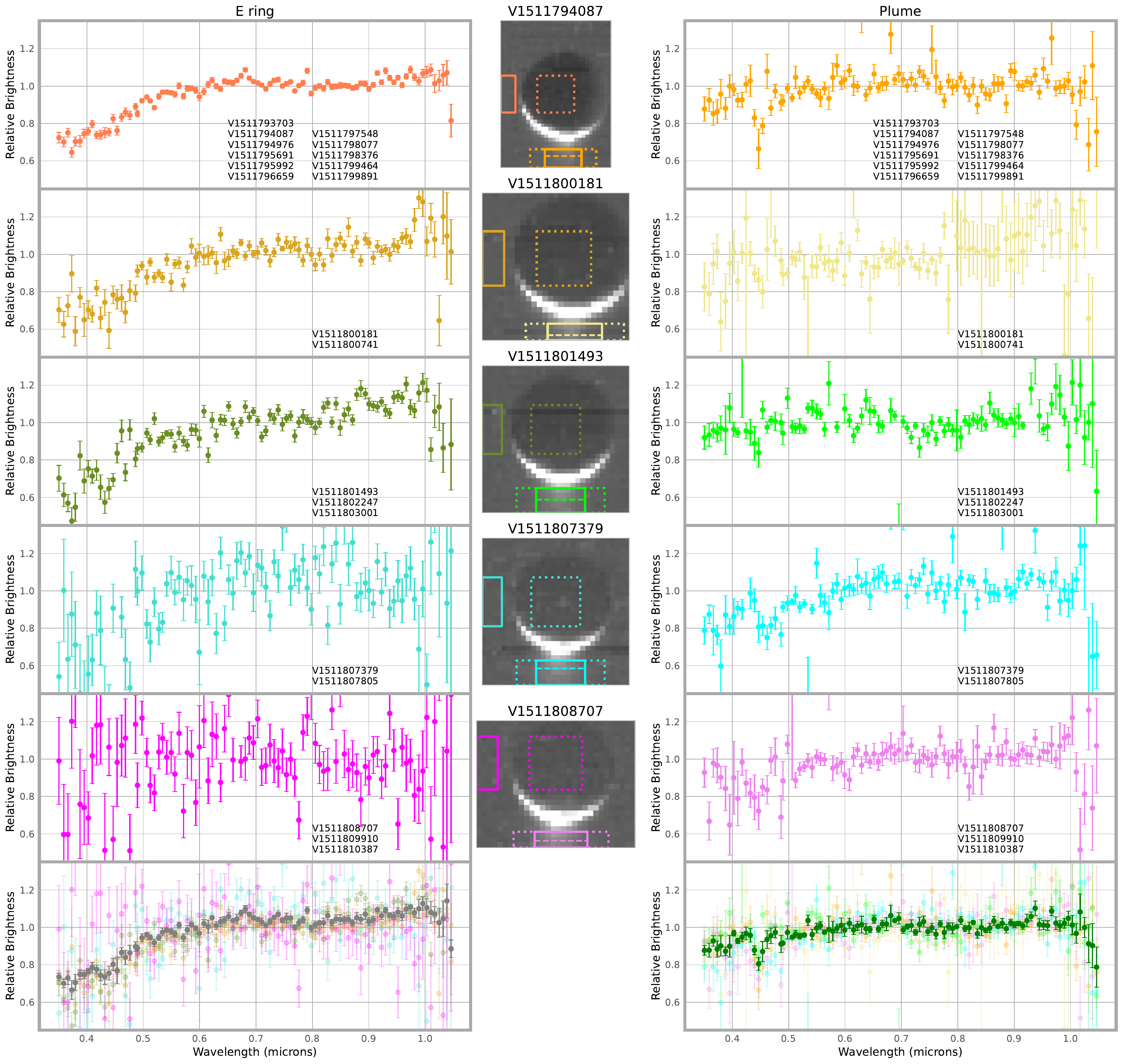}}
 \caption{Visible spectra of the Enceladus plume and E ring from VIMS. The central panels show representative images of Enceladus and its plume that correspond to the average brightness across all VIMS channels, and has been rotated so that the north pole of Enceladus points upwards. In all these images Enceladus is clearly visible as a dark disk against the background E ring  with a lit crescent in the south, and the plume can be seen emerging from the south pole. The panels on the left show the spectra of the E ring derived from the five sub-sets of the VIMS cubes, which are each derived from the average brightness in the region marked with the solid-edged box in the corresponding color in the corresponding image. The dotted edge box in the same color shows the region on the dark side of Enceladus used to define the background for these spectra. The right panels show the spectra of the plume derived using the data indicated by the solid box in the corresponding image, with the dashed line showing where the spectrum is evaluated and the dotted regions on either side determining where the background spectra was measured. The bottom panels show the average of the spectra derived from the five spectra above.}
 \label{plumevims}
 \end{figure}

Table~\ref{vimstab} lists the specific cubes considered for this analysis, which are the subset of cubes considered by \citeA{Hedman2009} with VIS exposure times longer than 7 seconds. These cubes were calibrated using the RC19 version of the VIMS calibration pipeline \cite{Brown2004, RC19}. These cubes belong to five different observation sequences which we designate by the numbers used in the original observation names (003, 010, 001, 011 and 007).  Note that the VIS cubes exhibit substantial brightness offsets between adjacent columns (rows in the rotated versions of the images shown in Figure~\ref{plumevims}, {the horizontal bands in some of these images are residual offsets left over after initial processing to remove these features}), which impacts how we process these data in order to extract spectra of the E ring and plume. 

Due to the E ring being a nearly constant background in the image, obtaining its spectrum is relatively straightforward.  For each cube, we take the image at each wavelength and select out a set of 10 columns that contain both the dark side of Enceladus and the background E ring, which are listed in Table~\ref{vimstab2}. In each of the selected columns, the E ring can be found in rows 0-3, while the dark side of Enceladus occupies rows 10-19. We therefore estimate the E-ring signal as the difference in the average signal between these two sets of pixels after applying median filters to remove cosmic rays and other instrumental artifacts. We also estimate the uncertainty on the E-ring signal based on the standard deviation of the pixels on Enceladus' dark side divided by the square root of the number of pixels on E-ring minus one. 

Initial inspection of these E-ring spectra revealed that they had consistent shapes, but that the overall brightness of the ring declined over the course of the observations. This trend can also be seen in the images in Figure~\ref{plumevims}, and most likely arises because of changes in the projected amount of E-ring material along the spacecraft's line-of-sight. Note that the sub-spacecraft longitude on Enceladus steadily increased from 100$^\circ$ to 150$^\circ$ over the course of the observation, so near the start of the observation the spacecraft is looking nearly along Enceladus' orbit (i.e. down the core of the E ring), whereas near the end it is looking more towards Saturn, so the line of sight passes through less E-ring material. 

Since this initial study is primarily concerned with the shape of the E-ring spectrum rather than its overall brightness, we opted to normalize each spectrum so that we could combine data from different images and thereby obtain high signal-to-noise spectra to compare with the ISS data. We therefore normalized each spectrum from each column in each image by the median brightness across all wavelengths, and then for each observation we computed the weighted average of the normalized spectra from the selected 10 rows and all the relevant data cubes. The uncertainty on this estimate is the corresponding weighted error on the mean, scaled by the square root of the reduced $\chi^2$ (typically between 1 and 2), which accounts for the excess scatter among the different spectra. The resulting five spectra are shown in the left panels of Figure~\ref{plumevims}. The signal-to-noise clearly decreases over the course of the observations, but there is a consistent  downturn at short wavelengths within all the observations with sufficient signal-to-noise to detect it. Finally, we take the weighted average of the five of these spectra to get an average normalized E-ring spectra, which is shown in the bottom panel of Figure~\ref{plumevims} and provided in tabular form in Table~\ref{vimsspectab}. Again, we scale the errors by the square root of the reduced $\chi^2$ (here of order 3) to account for any potential time variability in the plume spectra, although none were obvious in these data.

Extracting spectra of the plume is more challenging than the E ring because the plume is located close to Enceladus' lit limb and because it is a more compact source of light, whose brightness steadily decreases with distance from the limb. In this context, it is important to note that the VIMS-VIS spectrometer exhibits a spectral tilt that causes the images to systematically shift between columns as the wavelength increases \cite{McCord2004, Filacchione2007}. For these specific observations, this means that the same column in the image corresponds to higher altitudes above Enceladus' surface at longer wavelengths, which will introduce a spurious blue slope in the spectrum if not accounted for properly. 

Our procedure for extracting plume spectra starts by obtaining brightness spectra of the plume from each column at each wavelength in each cube. These spectra are derived by taking the average brightness in a set of 10 rows that contain the plume and subtracting the average of two sets of four rows on either side of the plume (The exact columns are provided in Table~\ref{vimstab2}). Again, median filters are applied prior to computing these averages to remove cosmic rays and instrumental artifacts. 

Due to the spectral tilt and the changing distance to Enceladus, we cannot simply average these individual spectra together to get a sensible spectra of the plume. Instead, we take all the data from each observation, and fit the brightness trend versus altitude at each wavelength in order to estimate the brightness of the plume at a fixed reference altitude. More specifically, we estimate the altitude probed  by each column $x$ in a particular wavelength channel $w$ (ranges from 0-95) using the following formula.
\begin{equation}
z=(x_c+0.0075*w-x)*(0.005/3)R-250 km
\end{equation}
where $x_c$ is column that should contain Enceladus' center (based on examination of the images; tabulated in Table~\ref{vimstab2}), and $R$ is the range to Enceladus (provided in Table~\ref{vimstab}). The factor of 0.0075$w$ empirically accounts for the shifts in pointing over wavelength (this correction was verified against an independent spectral tilt correction provided by G. Filacchione), and 0.005/3 is the pixel size in radians for the VIS channel in HIRES mode. Based on prior investigations of the VIMS plume observations \cite{Hedman2013, Sharma2023}, the brightness of the plume should decrease quasi-linearly with the parameter $Z$, which can be computed from the altitude $z$ using the following formula:
\begin{equation}
Z=\sqrt{\frac{z}{z+250 km}}.
\end{equation}
We confirmed the brightness data followed a reasonably linear trend and so fit the data within each observation at each wavelength obtained at $Z$ between 0.3 and 0.7 to a line in order to estimate the plume's brightness at $Z=0.5$ at all wavelengths for each observation, with an uncertainty derived from the scatter around the trend. 

The resulting spectra are shown in the right-hand panels in Figure~\ref{plumevims}.  Again, since we are more interested in the shape of the spectrum than the overall brightness, all of these spectra are normalized relative to their median value. In this case, the signal-to-noise of these spectra are more consistent among the five observations. In part this is because the changing viewing geometry should not affect the plume signal as much, and in part it is because the plume itself is increasing in brightness over the course of this observation \cite{Hedman2013}. These spectra also show evidence for a reduction in brightness at short wavelengths, albeit not as strong as that seen in the E ring. Again, since the differences in these normalized spectra are relatively subtle, we combine these spectra to obtain a single plume spectrum by taking the weighted average of the five spectra. Again, we scale the error bars by the square root of the reduced $\chi^2$ to account for the excess variations over time, and the resulting spectra are shown in the bottom right-hand panel of Figure~\ref{plumevims} and provided in Table~\ref{vimsspectab}.

\begin{table}
\caption{Average Normalized spectra of the Plume and the E-ring from VIS observations}
\label{vimsspectab}
\resizebox{!}{4in}{\begin{tabular}{|c|cc|cc||c|cc|cc|} \hline
Wavelength & Plume & Plume  & E ring  & E ring & Wavelength & Plume & Plume  & E ring  & E ring \\ 
(microns) & Signal  & Error & Signal & Error & (microns) & Signal  & Error & Signal & Error \\ \hline
0.3505&0.877&0.026&0.736&0.020&0.7029&0.987&0.019&1.043&0.011\\
0.3590&0.878&0.038&0.700&0.023&0.7100&1.004&0.014&1.030&0.019\\
0.3663&0.928&0.028&0.730&0.037&0.7173&1.010&0.031&1.011&0.010\\
0.3732&0.884&0.044&0.665&0.052&0.7248&0.973&0.015&1.032&0.015\\
0.3795&0.901&0.054&0.705&0.043&0.7320&1.016&0.039&1.054&0.009\\
0.3879&0.872&0.023&0.748&0.025&0.7393&1.031&0.029&1.070&0.010\\
0.3952&0.899&0.058&0.748&0.022&0.7468&0.978&0.020&1.038&0.013\\
0.4025&0.968&0.015&0.766&0.013&0.7540&0.998&0.025&1.050&0.026\\
0.4095&0.928&0.013&0.785&0.029&0.7613&0.964&0.026&1.051&0.004\\
0.4173&0.948&0.025&0.776&0.024&0.7687&1.019&0.017&1.000&0.014\\
0.4244&0.968&0.035&0.744&0.029&0.7760&0.972&0.032&1.026&0.018\\
0.4318&0.958&0.036&0.734&0.039&0.7833&1.005&0.019&1.042&0.017\\
0.4392&0.878&0.023&0.755&0.037&0.7907&0.995&0.029&1.093&0.018\\
0.4465&0.805&0.034&0.802&0.033&0.7979&1.004&0.008&0.995&0.017\\
0.4537&0.871&0.054&0.769&0.071&0.8052&1.037&0.021&1.015&0.019\\
0.4616&0.935&0.047&0.866&0.024&0.8126&0.994&0.016&1.025&0.005\\
0.4684&0.955&0.041&0.840&0.038&0.8199&0.998&0.031&1.024&0.024\\
0.4762&0.973&0.029&0.864&0.040&0.8272&1.026&0.011&1.062&0.023\\
0.4863&0.911&0.046&0.897&0.026&0.8346&0.967&0.025&1.027&0.016\\
0.4897&0.916&0.006&0.888&0.015&0.8419&0.999&0.042&1.045&0.021\\
0.4978&0.973&0.035&0.944&0.021&0.8492&0.984&0.012&1.042&0.026\\
0.5063&0.948&0.009&0.961&0.012&0.8566&0.997&0.025&1.041&0.018\\
0.5122&0.964&0.020&0.946&0.013&0.8639&1.034&0.024&1.042&0.021\\
0.5196&0.953&0.016&0.918&0.020&0.8712&1.005&0.009&1.025&0.016\\
0.5277&0.959&0.012&0.945&0.016&0.8786&1.022&0.027&1.051&0.029\\
0.5342&1.037&0.027&0.975&0.020&0.8859&0.992&0.023&1.043&0.025\\
0.5416&0.962&0.028&0.985&0.009&0.8939&1.025&0.018&1.071&0.027\\
0.5495&0.996&0.026&0.979&0.015&0.9003&1.015&0.016&1.030&0.021\\
0.5561&1.010&0.020&0.965&0.016&0.9079&1.020&0.013&1.042&0.017\\
0.5635&0.975&0.011&1.012&0.016&0.9152&1.009&0.020&1.095&0.007\\
0.5713&0.990&0.070&0.945&0.021&0.9225&1.037&0.033&1.062&0.020\\
0.5781&0.979&0.052&0.986&0.023&0.9298&1.058&0.024&1.092&0.013\\
0.5855&0.996&0.023&1.014&0.010&0.9371&1.102&0.039&1.065&0.013\\
0.5931&1.010&0.040&0.995&0.013&0.9445&1.016&0.030&1.061&0.019\\
0.5994&0.986&0.034&0.968&0.021&0.9518&1.012&0.024&1.081&0.024\\
0.6076&1.022&0.027&1.020&0.021&0.9591&1.009&0.026&1.061&0.025\\
0.6150&0.986&0.025&1.004&0.034&0.9664&1.087&0.024&1.098&0.025\\
0.6221&1.008&0.023&1.052&0.007&0.9738&1.023&0.034&1.110&0.008\\
0.6294&1.014&0.045&1.010&0.020&0.9810&1.008&0.027&1.079&0.024\\
0.6370&0.993&0.028&1.030&0.025&0.9888&1.022&0.026&1.100&0.050\\
0.6441&0.989&0.017&1.040&0.018&0.9959&1.049&0.038&1.128&0.037\\
0.6514&1.043&0.020&1.059&0.020&1.0029&1.012&0.052&1.105&0.049\\
0.6591&0.991&0.017&1.037&0.006&1.0100&0.968&0.053&1.080&0.058\\
0.6661&1.018&0.020&1.044&0.026&1.0170&1.083&0.096&1.071&0.040\\
0.6734&1.018&0.035&1.077&0.016&1.0247&0.999&0.053&1.025&0.078\\
0.6810&1.063&0.056&1.098&0.018&1.0319&0.913&0.102&1.046&0.095\\
0.6880&1.037&0.004&1.070&0.010&1.0387&0.903&0.118&1.142&0.090\\
0.6953&1.048&0.030&1.051&0.007&1.0460&0.787&0.108&0.886&0.047 \\
\hline
\end{tabular}}
\end{table}

\section{Results and Discussion}
\label{Results}

\begin{figure}
\resizebox{\textwidth}{!}{\includegraphics{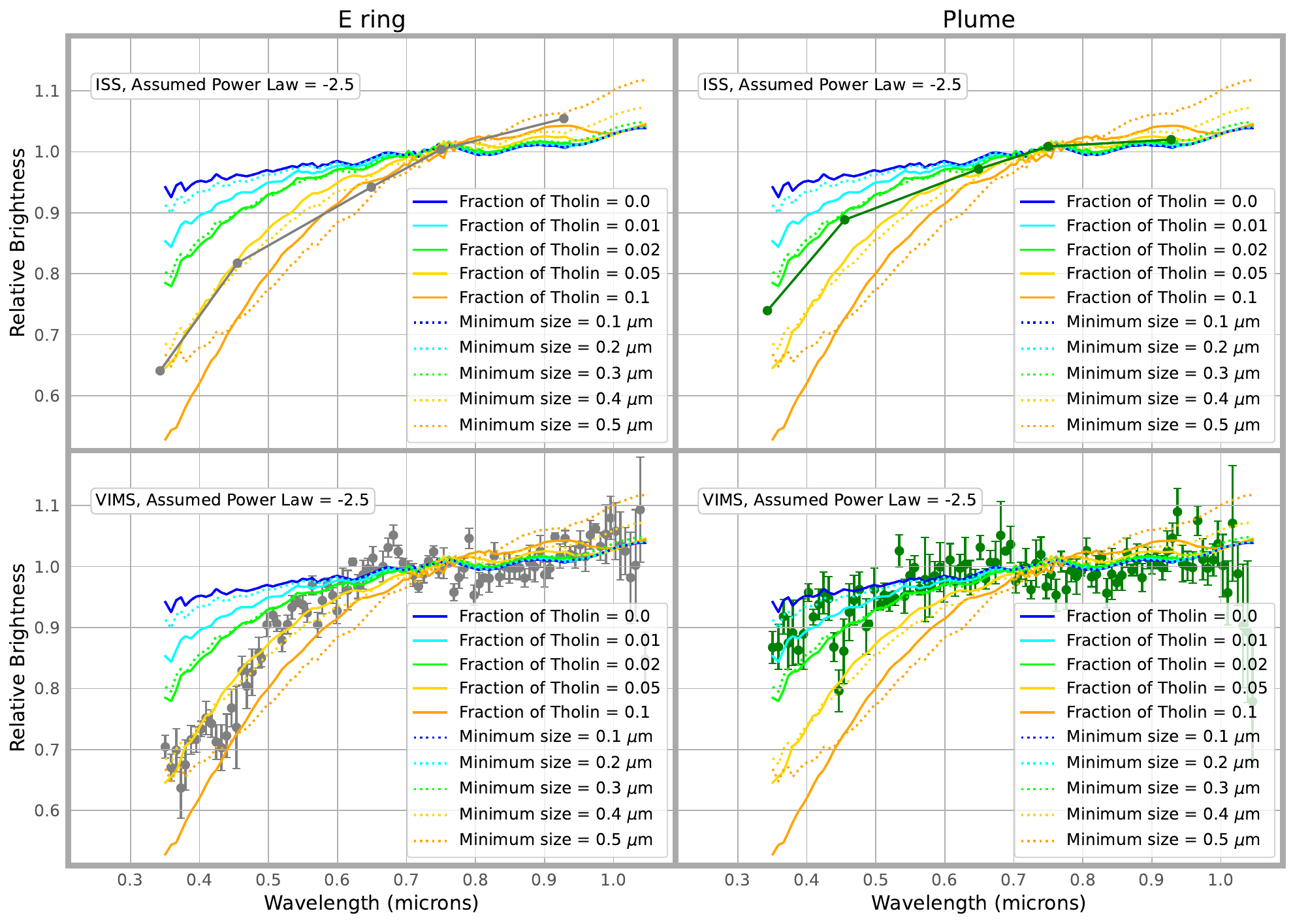}}
\caption{Summary of the visible spectra of the E ring and Enceladus Plume from Cassini ISS and VIMS observations. Each panel shows a normalized spectrum of a relevant dust population (normalized to have an average value of 1 between wavelengths of 0.6 $\mu$m and 0.95 $\mu$m. Overlaid on the observed spectra are Mie-theory predictions for various populations of particles. The solid lines correspond to size distributions that follow a power law with index -2.5 between 0.1 $\mu$m and 5.1 $\mu$m that are assumed to be composed of water ice mixed with different amounts of tholins using optical constants from \citeA{Warren2008} and \citeA{Baratta2015}. The dotted lines show predictions for pure ice particles following a power law with index -2.5 between the indicated minimum size and 5.1 $\mu$m.  Note these model spectra are not formal fits to the observations, and are instead sample calculations to highlight trends with composition and particle size cut-offs.}
\label{specsum}
\end{figure}

Figure~\ref{specsum} provides an overview of the E-ring and Enceladus plume spectra derived from the ISS and VIMS observations. To facilitate comparisons, all spectra have been normalized to have an average value of unity between 0.6 $\mu$m and 0.95 $\mu$m. These spectra do reveal some differences and discrepancies that might reflect residual instrumental or data-processing artifacts. First of all, the spectral slope between 0.6 $\mu$m and 1.0 $\mu$m measured by ISS is  redder than the slope measured by VIMS for both the E-ring and the plume. Most notably, the normalized brightness values for the ISS RED filter at 0.65 $\mu$m is roughly 5\% below the corresponding values in the VIMS spectra. This long-wavelength discrepancy most likely reflects residual errors in the relative calibration of the two instruments. Mie-theory calculations indicate that these changes cannot be explained by the slightly higher phase angle of the ISS observations, and while the observations do occur at very different Enceladus orbital phases, the fact that the difference is seen in both the E ring and the plume argues against this being the explanation for the discrepancies at lengths longer than 0.6 $\mu$m. Similarly, the fact that both the E ring and plume spectra from VIMS show a $\sim$5\% peak at around 0.67 $\mu$m and a broad $\sim$2\% dip between 0.7 $\mu$m  and 0.9 $\mu$m suggests that these features are due to small residual differences in the calibration, and so we will not consider these features further here.

The situation at shorter wavelengths is more complicated. For both the E ring and the plume, the relative brightness measured by ISS with the BL1 filter around 0.45 $\mu$m  is within a few percent of the average value for the corresponding wavelengths observed by VIMS, although the comparison is complicated by the relatively steep spectral slope in the E-ring spectrum and the dip in the plume spectrum at these wavelengths, which will be discussed further below. However, the relative brightness of the plume measured by ISS with the UV3 filter at 0.35 $\mu$m is roughly 10\% lower than the relative brightness measured by  VIMS, while the ISS value is only 5\% lower than the VIMS value for the E ring. The $\sim$5\% E-ring difference between the two instruments is comparable to that seen at longer wavelengths, and since the absolute calibration is less certain at these wavelengths \cite{RC19, Knowles2020}, this discrepancy is most likely due to residual calibration uncertainties. However, the larger difference between the ISS and VIMS data for the plume could potentially involve a real difference in the plume's spectral properties. These two observations were obtained at very different orbital phases, {and recent models of Enceladus' tectonics suggest that the plumes' particle composition could vary as the stresses experienced by the South Polar Terrain change \cite{Soucek2024}. Given that the ISS data already indicate different sources could have different spectral properties, we expect that future examinations of a broader range of plume observations will clarify how the plume's spectral  properties vary with orbital phase. }

Interestingly, the VIMS plume spectrum appears to have a $\sim$10\% dip centered around 0.45 $\mu$m. At the moment, we do not have a full explanation of this feature. Comparing the spectra from the different parts of the VIMS observation in Figure~\ref{plumevims}, this feature is not restricted to a particular subset of the data, and visual inspection of the cubes did not reveal obvious artifacts at those wavelengths. A weaker dip at similar wavelengths can also be seen in the E-ring spectra. However, it is unlikely that this particular feature is a calibration artifact because this feature appears to be centered around slightly different wavelengths in the two spectra, and is not obviously present in the surface spectra shown in Figure~\ref{surfspec}. This feature may therefore be a real spectral feature in the plume. Since it is only marginally significant, and we are not aware of any obvious candidate for what might create such a band, we will not attempt to interpret this feature further here.

Importantly, all the spectra do show evidence for a change in spectral slope around 0.5-0.6 $\mu$m that appears to be consistent with the signature of  a UV absorber (like tholins) shown in Figure~\ref{theory}. In principle, we could quantitatively constrain the composition of the plume by comparing these spectra with predictions from Mie theory. {However, in practice contaminants and features in the size distribution can have similar effects on the observed spectrum (see Figure~\ref{theory}), and so there is unlikely to be a unique best-fit model for any individual spectrum. Indeed, initial studies of these spectra demonstrated that quantitative estimates of the particles' composition depended on assumptions about both the particle size distribution and the properties of the contaminant, and so obtaining robust constraints on particle properties would require exploring a relatively large parameter space. Furthermore, the easiest way to distinguish the effects of the particles' composition and size distribution is by considering spectra obtained at multiple phase angles. These sorts of analyses are beyond the scope of this initial study. Instead, we offer preliminary comparisons  that provide rough estimates of the tholin fraction in order to guide future studies.}

Specifically, we consider the same Effective Medium Theory mixtures of water and tholin used previously, and consider simple power-law size distributions with abrupt cut-offs. The corresponding Mie spectra are computed using the {\tt SF\_SD} function from the  PyMieScatt package \cite{Sumlin2025}, and are overlaid on the various observed spectra in Figure~\ref{specsum}. To simplify things, we compare all four observed spectra to the same set of models, all of which assume the particles are composed primarily of water ice with  \citeA{Warren2008} optical constants and follow a size distribution with the same {differential} power law index of -2.5 and the same maximum particle size of 5.1 $\mu$m. The solid lines show the expected spectra of particle populations where the power-law size distribution extends down to 0.1 $\mu$m, and have different amounts of tholins with \citeA{Baratta2015} optical constants. The dotted lines show the predicted spectra of pure-ice particle populations with the indicated minimum particle sizes. These curves demonstrate the challenges with distinguishing the effects of a non-ice contaminant from a deficit of sub-micron particles, but are still sufficient to place some rough constraints the composition and size distribution of the E ring and plume.


The plume spectra are both consistent with tholin fractions around 1-2\% or a particle size distribution lacking particles smaller than 0.2-0.3 $\mu$m. The required cut-off in the plume's particle size distribution is probably incompatible with the in-situ evidence that the particle size distribution follows a power-law down to  a few nanometers \cite{Dong2015}, but the required tholin fraction could be consistent with the fraction of grains containing high-mass organic compounds reported by CDA \cite{Postberg2018, Nolle2024}. Thus it is reasonable to conclude that the spectral signature in the plume reflects the plume particles' chemical composition. {Also, while we cannot completely rule out the possibility that this signature is due to the plume particles containing a smaller fraction (0.1-0.2\%) of iron compounds like hematite, there are currently no explicit predictions or constraints for how much of this material could be incorporated into the plume particles. Hence, at this point it would appear most likely that this feature is due to the plume particles containing 1-2\% complex organic compounds.} 

The E ring spectra show a much stronger change in spectral slope than the plume, which could indicate a tholin fraction of around  5\% or a minimum particle size of around 0.4 $\mu$m. A higher tholin fraction in the E ring could be consistent with recent CDA measurements of the plume, which indicate that the lower-velocity diffuse sources might be more salt-rich than organic rich \cite{Ershova2024}, resulting in a larger fraction of organic-rich particles reaching the E ring.  It is also possible that the tholin-like signature develops over time as the particles age and erode \cite{Nolle2024}. {Recall that the stronger UV absorption on moons far from Enceladus' orbit has been attributed to higher fluxes of energetic charged particles in those regions \cite{Hendrix2018b}, so perhaps exposure to energetic electrons also enabled E-ring particles to develop stronger UV absorptions over time.  However, it is also possible that the E-ring may have a smaller fraction of sub-micron particles than the Enceladus plume.  As mentioned above, published in-situ measurements of the particle size distribution in the E ring \cite{Ye2014, Ye2016, Ye2018} provide only limited information about the size distribution of particles smaller than 1 $\mu$m, and earlier spectrophotometric data suggest that the E ring may have few sub-micron particles \cite{Showalter1991}. Numerical simulations also indicate that such small particles have short lifetimes within the E ring \cite{Juhasz2007}.}  The E-ring's stronger UV absorption therefore could represent a combination of signals reflecting changes in both the particles' composition and their size distribution. 

The spectra presented here provide enough information to demonstrate that the high-phase spectra obtained by ISS and VIMS contain a spectral feature that encodes information about the plume's and E-ring's composition and/or size distribution. However, these data alone are insufficient to detangle the effects of particle composition and size distribution for these systems. Additional work will therefore be needed to extract robust information about the composition and size distribution of the plume and E ring. This will involve not only cross-comparisons of data obtained with different instruments, but also more comprehensive analyses of the spectral data over a wide range of lighting and viewing geometries. Such studies will not only clarify the average particle properties of the plume and the E ring, but should also provide new insights into how these properties vary with time and space. In particular, the above-mentioned variations in the plume's spectral properties with location (seen in the ISS data) and with time (in the VIMS data) would reveal whether different sources are more or less likely to erupt organic-rich materials under different conditions.

\section*{Open Research Section}
The raw ISS and VIMS data and calibration pipelines are available via the Planetary Data System. The calibrated images and spectral cubes, as well as the software used to generate the spectra and plots shown in this paper, are available via github \\ {\tt https://github.com/mmhedman/Enceladus\_Plume\_Color} and via \citeA{Hedman26}
\section*{Conflict of Interest disclosure}
The authors declare there are no conflicts of interest for this manuscript.

\acknowledgments
This work was partially supported by NASA Grant 80NSSC24K0421. We thank the members of the Cassini Mission, Imaging and VIMS Teams for obtaining the data used in this study. We thank G. Filacchione and M. Ciarniello for providing an independent check on our spectral tilt corrections for the VIS VIMS data, and thank R. Cartwright and F. Postberg for helpful discussions. We also thank the reviewers for their helpful comments on an earlier version of this manuscript.


\end{document}